\documentclass[preprint,12pt,authoryear]{elsarticle}

\usepackage{amssymb}

\journal{Planetary and Space Science}

\begin{document}

\begin{frontmatter}



\title{Searching sub-stellar objects in DR1-TGAS, effectiveness and efficiency of Gaias' astrometry}


\author{P.A.~Cuartas-Restrepo, O.A. S\'anchez-Hern\'andez, M. Medina-M}

\address{FACom - SEAP - Instituto de F\'{\i}sica - FCEN, Universidad de
  Antioquia, \\ Calle 70 No. 52-21, Medell\'{\i}n, Colombia.}

\begin{abstract}
We used 1,477,047 data from DR1-TGAS, in order to analyse the minimum
requirements of accuracy, necessary to detect sub-stellar objects in
the astrometric measurements of Gaia. We found that the first set of
data (DR1) does not have enough accuracy, so sub-stellar objects can
not be easily detected. Barely, it would be possible to detect jovian
and higher mass objects, with orbital periods over 5 years. We made
the calculations of the minimum values of the astrometric angle
produced by an orbiting sub-stellar object using a range of different
masses. We estimate the efficiency and effectiveness of the DR1-TGAS
data in order to detect sub-stellar objects and the minimum accuracy
that Gaia would be required to detect these objects using the datasets
that the mission will release in the near future.
\end{abstract}


\end{frontmatter}


\section{Introduction}
\label{Section:Intro}

Only one exoplanet has been discovered by the method of
astrometry\footnote{See in: http://exoplanet.eu/catalogue/}. It is a
jovian giant known as HD 176051 b, discovered in 2010
\citep{Muter10}. In addition to this, there is no other record in the
literature of new exoplanets discovered using astrometric data.

Some works in the past have addressed the possibility of detecting
planets using the astrometric measurements, especially based on those
made by \textit{Hipparcos} mission. Prior to Gaia, the
\textit{Hipparcos} mission was the more precise source of astrometric
measurements \citep{Perryman08}. Unfortunately the measures 
of \textit{Hipparcos} are not sufficiently
accurate, and should be supported by other measurements made both in
Earth, with the Multichannel Astrometric Photometer MAP
\citep{Gatewood97}, as well as with measures made by space missions
such the Full-Sky Astrometric Mapping Explorer FAME \citep{Horner99}
or the Space Interferometry Mission SIM \citep{Shao99}.  Sadly as we
all know, both missions were cancelled by NASA, first FAME was cancelled
in 2002, and finally SIM was cancelled in 2010.

\citet{Gatewood01} work combines observations of \textit{Hipparcos}
and MAP to analyse the \textit{$\rho$ Coronae Borealis} system. They
found that what appeared to be a planetary companion, it was actually
a red dwarf. The accuracy of the astrometric measurement made with MAP
confirmed that the mass of the object should be at least 100 times
greater than the mass of the sub-stellar companion initially reported
by \citet{Noyes97}, which was measured by radial velocity and was of
the order of 1.1 $M_J$.  In their work, \citet {Gatewood01} shows how
the measurement of the mass made by astrometry can be many times more
precise than that made by the radial velocity. One main conclusion of
their work is that only massive objects can actually produce a
measurable astrometry signal.

In the work of \citet{Han01}, the \textit{Hipparcos} data for 30 stars
with radial velocity periodic variations was reduced.  Their
preliminary results were proposed as a guide for the selection of
observational objectives for astrometry projects. Their results
compare the masses of sub-stellar companions in some systems
previously analysed by radial velocity such as \textit{Ups And, HD
  10687, 70 Vir} and \textit{47 UMa}. An important conclusion of their
work was the need for astrometric instruments of greater precision in
a range of 1-2 orders of magnitude higher than \textit{Hipparcos}. On
the other hand, the error in the measure of the mass of sub-stellar
companions should be associated with small angles of inclination,
which requires better astrometric measurements.

\begin{figure}
\centering
\includegraphics[width=0.6\columnwidth]{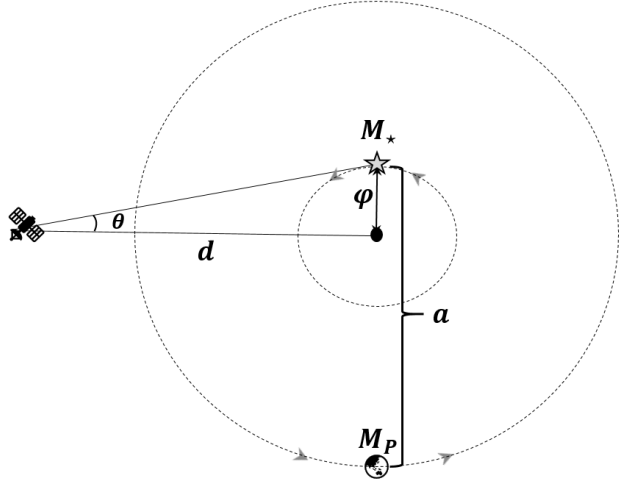}
\caption{The astrometric angle $\theta$.}
\label{Figure:AstroSignal}
\end{figure}

\textit{Hipparcos}' astrometric data were also used to estimate the
inclination of some planetary systems observed by radial velocity in
the work of \citet{Pourbaix01b}. A fundamental conclusion of this work
is that instruments with a precision of at least 100 $\mu as$ are
required, otherwise, astrometry techniques will not be able to measure
the mass of sub-stellar companions. The work of \cite{Pourbaix01a} 
concludes on the same terms that the
\textit{Hipparcos} data is not enough to show if companion candidates
could be planets or bodies of stellar nature.

The Gaia mission was launched in December 2013 with the aim of
determining the accurate position and distance of more than 1 billion
stars in the galaxy \citep{Prusti16}. For almost three years 
astronomers and planetary scientists around the
world were waiting for Gaia to reveal its first set of data,
especially the astrometry data, in order to begin the search for
signals that allowed us to infer the presence of extrasolar
planets. Finally in September 2016, the mission revealed the first set
of data \citep{Brown16}. This first package (DR1) contains a total of
1,142,679,769 sources, and is divided into three main groups of data:
1) 93,635 shared data with \textit{Hipparcos}, 2) 1,963,415 data
shared with Tycho-2 and \textit{Hipparcos}, and 3) an additional
package of 1,140,622,719 secondary data \citep{Brown16}. A detailed
description of the astrometry data can be found in \cite{Lindegren16}.

Before the first observational results were known, calculations had
been made on the real possibilities of detecting exoplanets using the
astrometry data from Gaia \citep{Perryman14}. Based on the
signal-noise ratio, \cite{Perryman14} predicted a very high number of
possible detections. At least 21,000 giant planets with masses between
1.0-15.0 $M_J$ with long period could be discovered around stars at
distances of up to about $\sim500$ pc during the nominal period of the
mission which is five years. Even the work of \cite{Perryman14}
estimates that between 1,000-1,500 planets could be detected around M
dwarfs within a 100 pc distance. The total number of planets at the
end of the mission, in about 10 years, could reach 70,000!

We analyse the actual possibilities of finding sub-stellarmobjects 
in DR1-TGAS and which is the actual effectiveness
and efficiency of the astrometric measurements is this first release. 
This work is organized as follows. The section \ref{Section:Astro} is
dedicated to explain how the astrometry method works. In section
\ref{Section:DR1} we explain how the DR1-TGAS data is organized and
the analysis we have done to it. The section \ref{sec:EE} shows our
results related to the effectiveness of the astrometric data. Finally
in section \ref{sec:DisCon} we discuss our results and the real
possibilities of finding sub-stellar objects using DR1-TGAS data.

\section{Using astrometry to find planets}
\label{Section:Astro}

In a two body system, like a star-planet system, both orbits their 
common centre of mass. The star is displaced from the centre of mass
by a distance $\varphi$. Viewed from the Earth, this displacement is
observed as an angular distance $\theta$ (see Figure
\ref{Figure:AstroSignal}). The angle $\theta$ is equivalent to the
apparent movement of the star over the plane of the sky. This angle
can be measured by comparing the changes in the instantaneous position
of the star across the time, measured as an astrometric signal. If the
measure has the sufficient precision, it is possible to infer the
existence of a low-mass object orbiting around the star, i.e. a
planet.

From the Newtons' laws, we can determine the measurement of the
astrometric angle $\theta$ as \citep{Quirrenbach10}:

\begin{equation} 
\label{Equation:AstroSignal}
\theta = \left({\frac{G}{4 \pi^2}}\right)^\frac{1}{3} \left(\frac{M_p}{M_\oplus}\right)
\left(\frac{M_\star}{M_\odot}\right)^{-\frac{2}{3}}
\left(\frac{P}{yr}\right)^{\frac{2}{3}}\left(\frac{d}{pc}\right)^{-1}
\end{equation}

Here $G$ is the Cavendish constant, $M_P$ is the mass of the secondary 
object that disturbs the star of mass $M_\star$, $P$ corresponds to the
orbital period of the planet and $d$ is the distance between the
measuring instrument (Gaia in our case) and the extrasolar system. 
This expression for the astrometric angle $\theta$ is independent of
the inclination of the orbital plane. This enhances the method of
astrometry for the determination of the secondary object mass $M_P$ with
respect to other methods, like radial velocity, because it allows us
to obtain the precise mass of the object. Usually angle $\theta$ is
expressed in microarcsec units ($\mu as$) which are also an indication
of the level of accuracy required on the measuring instrument.

\begin{table}
\centering
\caption{Astrometric angle in $\mu as$ calculated for sub-stellar objects of different 
  masses, with a period of P=5 yr, orbiting around low-mass stars, at a distance of 100 pc.}
\vspace{0.4cm}  
\label{Table-Parameters}
\begin{tabular}{cccc}
   \hline
    $M_{\star} \ (M_{\odot})$ & $1.0 \ M_J$     & 10.0 $M_J$    & 20.0 $M_J$ \\
   \hline
   \hline
   0.1      & 129.40   & 1294.03  & 2588.06 \\
   \hline
   0.4      & 51.35    & 513.54   & 1027.07 \\
   \hline
   0.8      & 32.35    & 323.51   & 647.02 \\
   \hline
   1.0      & 27.88    & 278.79   & 557.58 \\
   \hline
\end{tabular}
\end{table}

\section{Gaia Data Release 1 - DR1}
\label{Section:DR1}

\subsection{Number of objects and their distances}
\label{Subsection:DNO}

Our analysis starts with the $2,057,050$ objects included in DR1-TGAS, 
and shared with \textit{Hipparcos} and \textit{Tycho-2} catalogues. Of this
total, we discard the objects with negative parallaxes
($\widetilde{\omega} < 0$), and with signal-noise ratio
$\widetilde{\omega}/\sigma_{\widetilde\omega}<3$, with which the size
of our dataset was reduced to $1,477,047$ objects.  A similar
debugging was done by \cite{McDonald2017} in the determination of the
luminosities of this same set of objects. \cite{Brown16} used a more
demanding signal-noise ratio, $\widetilde{\omega}/\sigma_{\widetilde
  \omega}<5$.

\begin{figure}
\centering
\includegraphics[width=0.8\textwidth]{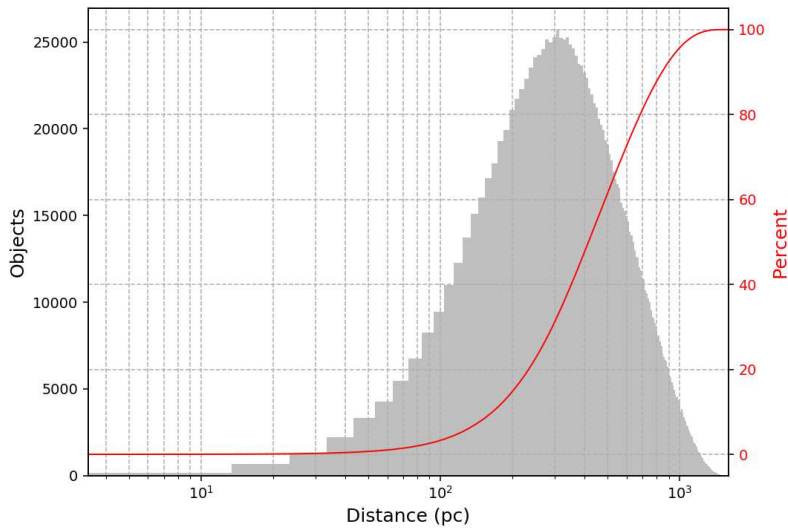}
\caption{Histogram of frequency of the number of objects and their distance 
in pc. Almost $\sim 60\%$ of the objects are closer than 500 pc.}
\label{Figure:Hist-Distance}
\end{figure}

Figure \ref{Figure:Hist-Distance} shows the distribution of distances
for our dataset. It is observed that almost $95\%$ of the objects are
concentrated in distances up to 1 $kpc$. $60\%$ of the data are closer
than 500 $pc$.  While the astrometric angle $\theta$ decreases with
the distance $d$, it is constituted as a border condition for the
detectability of sub-stellar objects.

\begin{figure}
\centering
\includegraphics[width=0.8\textwidth]{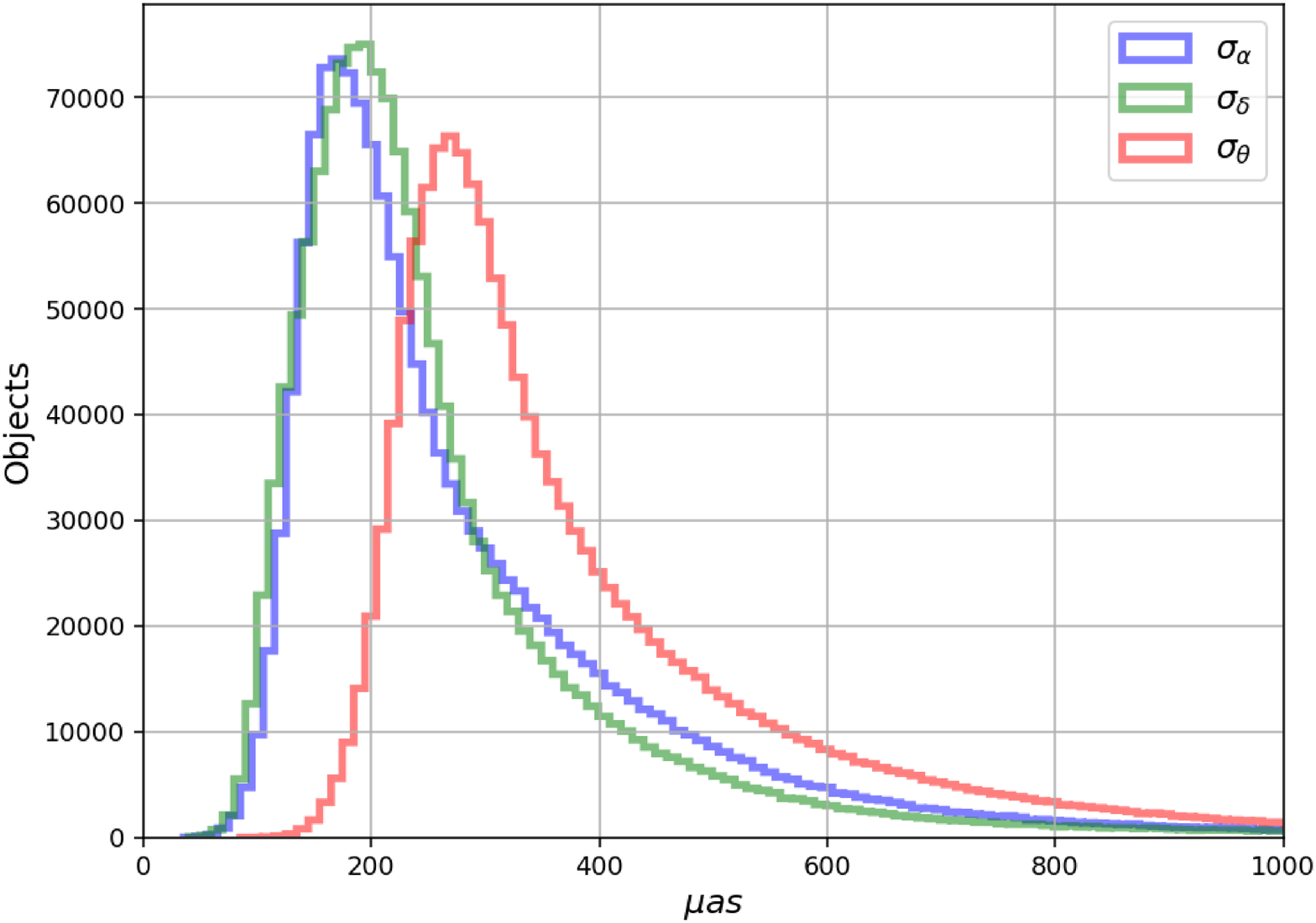}
\caption{Distribution of the uncertainty $\sigma_{\alpha}$ 
  (right ascension), $\sigma_{\delta}$ (declination) and $\sigma_{\theta}$ 
  (astrometric angle), of the Gaia-\textit{Hipparcos} 1,477,047 objects.}
\label{Figure:RA-Dec-Errors}
\end{figure} 

\subsection{Error in the position of an object}
\label{Subsection:ETP}

It is also relevant the determination of uncertainties
$\sigma_{\alpha}$ and $\sigma_{\delta}$ in the measurements of the
right ascension and declination of the $1,477,047$ objects. The
accuracy in the measurement of $\theta$ depends on the uncertainty in
$\alpha$ and $\delta$ measurements. Figure \ref{Figure:RA-Dec-Errors} 
shows the distribution of both
uncertainties, $\sigma_{\alpha}$ (red line) and $\sigma_{\delta}$
(green line). The error in both cases is centred around 200 $\mu
as$ and 80\% of data have error less than 300 $\mu as$.

Figure \ref{Figure:AstroSignal} shows the astrometric angle subtending
the semimajor axis of the orbit of the star around the centre of mass,
therefore the error on the astrometric angle measurement is determined
by the error on the position of the star at each side of the semimajor
axis.  This means that the astrometric angle measurement should be, at
least, greater than the uncertainties in the location of the star to
avoid false positive in exoplanets searching. Normally the measured
signal should be, at least, from three to five times the error
\citep{Sozzetti2014}.

On the other hand, the distribution of the total uncertainty,
$\sigma_\theta$ is the uncertainty on the location of the objects on
the plane of the sky, as a result of the combination of right
ascension and declination uncertainties. If we consider the errors in
spherical coordinates, then the uncertainty $\sigma_\theta$,
corresponds to a distance between the points $(\alpha+\sigma_{\alpha},
\delta+\sigma_\delta)$ and $(\alpha-\sigma_{\alpha},
\delta-\sigma_\delta)$.  The observed object is located at any point
within the area forming this spherical square. In a good approximation, 
the uncertainty on the astrometric angle
$\sigma_\theta$ which determinates the length of an arc in spherical
coordinates is calculated as follows:

\begin{equation}
\label{Equation:Error}
\sigma_\theta = \sqrt{\left(X_1-X_2\right)^2 + \left(Y_1-Y_2\right)^2},
\end{equation} \\

It is evident how the error in astrometric angle accumulates around
270 $\mu as$. More than 80\% of the objects have astrometric angle
error less than 600 $\mu as$.

\subsection{Mass and luminosity of Gaias' stars}
\label{Subsection:Mass ans luminosity}

Based on the parallax measurements reported in DR1-TGAS and the
measurements of the G-band magnitude supplied by Gaia, we proceeded to
estimate the stellar masses $M_G$ of each of the objects. For this, we
calculate the luminosity $L_G$, which corresponds to an approximation
of the actual luminosity, and based on a mass-luminosity relationship,
we find finally $M_G$.

These calculations do not take into account the extinction in the
measure of the brightness of the stars caused by the interstellar
medium. Neither correction of the absolute magnitude is included, from
which the bolometric magnitude is obtained.  \cite{McDonald2017}
concluded in his work that the extinction in most of the objects of
DR1-TGAS dataset is very low. On the other hand, \cite{Brown16} does
not make corrections for the magnitude in the G-band. The work of
\cite{Jordi10} did correction tests for the bolometric magnitude, but
these corrections that are less $\sim 0.15$ magnitudes, which is
negligible for our purpose.

\begin{figure}
\centering
\includegraphics[width=0.8\textwidth]{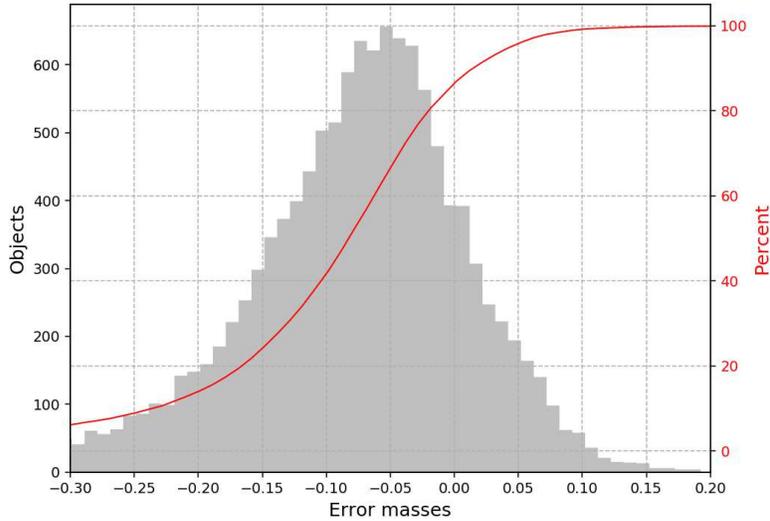}
\caption{Error histogram for masses of 12,533 stars of our dataset ($M_G$) versus 
masses calculated by \citet{Allende99} ($M_A$).}
\label{Figure-Error masses}
\end{figure}

Additionally, we compared our results with those published by
\cite{Allende99}, who calculated the stellar parameters, including the 
mass, for 17,219 stars in the Hipparcos catalogue within a distance of 
100 pc from the Sun. For comparison of the results we take the common data 
to both catalogs, a total of 12,533 stars. We discard those stars not 
included in our dataset. Figure \ref{Figure-Error masses} shows that 
the 80\% of the stars common to both datasets (12,533 stars), 
have errors $1-M_G/M_A$ concentrated between -15\% and 5\%.

\begin{figure}
\centering
\includegraphics[width=0.8\textwidth]{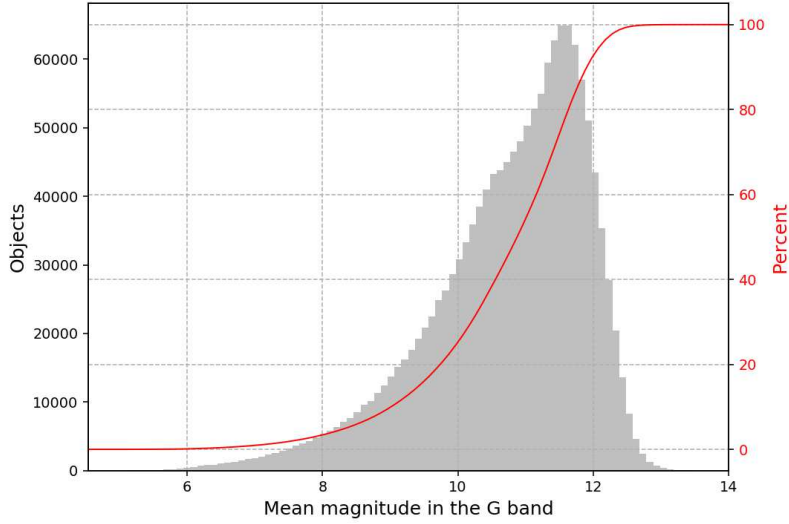}
\caption{Histogram of frequency of the number of objects versus G-Band magnitude. 
All stars of our dataset have maximum bright to 13.}
\label{Figure-Hist-Magnitudes}
\end{figure}

\begin{figure}
\centering
\includegraphics[width=0.8\textwidth]{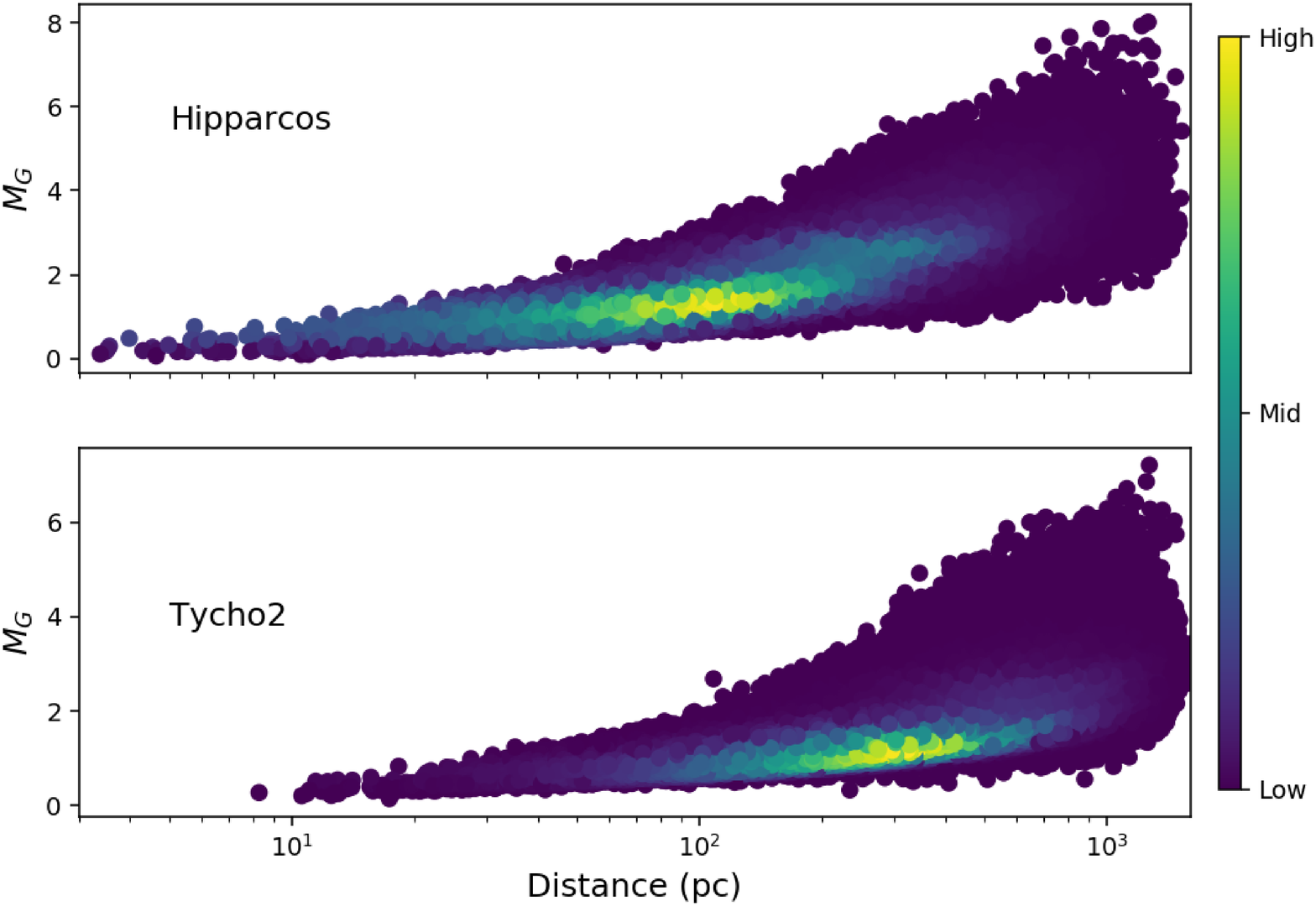}
\caption{Scatter of masses $M_G$ versus distances of our dataset for each catalogue. 
\textit{Up}: Hipparcos catalogue. \textit{Down}: Tycho catalogue.}
\label{Figure-Scatter-Distances-Vs-Masses}
\end{figure}

In conclusion, we find that the difference in the results for the stellar 
masses are negligible for the purposes of our analysis. Figure 
\ref{Figure-Hist-Magnitudes} shows the histogram of magnitudes 
in G-band, which shows that the stars in our catalogue are between 
6 and 13 magnitudes. Figure \ref{Figure-Scatter-Distances-Vs-Masses} 
shows a dispersion diagram of the Gaias' stellar masses $M_G$ and their 
distances $d$ in pc.

\section{Effectiveness and efficiency in DR1-TGAS data}
\label{sec:EE}

Based on DR1-TGAS data, and in the stellar masses estimated in the
previous section, we can calculate the minimum orbital period of the
star around the barycentre of the system caused by the presence of a
sub-stellar object. We will assume a minimum mass of the sub-stellar 
objetc $M_{obj}$, and a minimum astrometric angle $\theta$ equal to 
a signal-noise ratio $SNR \geq 3$, this is to ensure reliability in 
the detectability of the sub-stellar object.

\begin{equation}
\label{Equation:SNR}
SNR = \frac{\theta}{\sigma_\theta} \geq 3.
\end{equation}

Hence, the minimum value of $\theta$ is,

\begin{equation}
\label{Equation:Theta-Min}
\theta_{min} = 3 \sigma_{\theta},
\end{equation}

therefore, from Eq. \ref{Equation:AstroSignal} we will obtain the
minimum detection period,

\begin{equation}
\label{Equation:Period}
P_{min} \approx 1.27 \times 10^{-3} \ M_G \Biggl(\frac{\sigma_\theta d}{M_p}\Biggr)^{3/2} \ yr.
\end{equation}


\begin{figure}
\centering \includegraphics[width=0.8\textwidth]{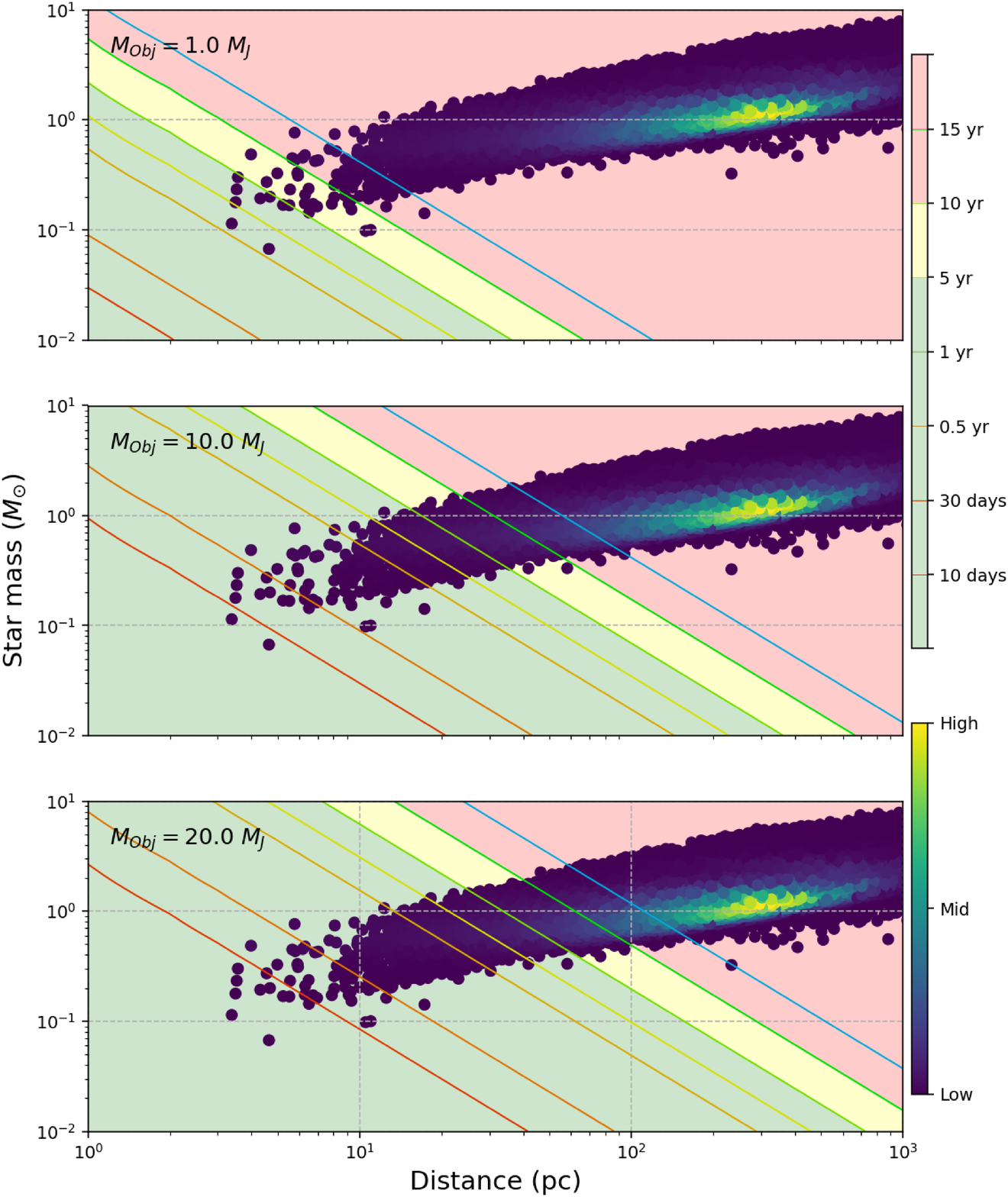}
\caption{Minimum period for sub stellar objects detection in function of the distance (pc) 
and of the $M_G$ for all stars of our dataset. Each point represents a particular object 
of our dataset of 1,477,047 stars according to its mass $M_G$ and the distance it is 
located. The contour lines, referenced with the upper right colour bar, indicate the 
minimum orbital period that the system of each particular star must have in order for 
a planet of mass $M_{obj}$ to be detectable. The lower right colour bar indicates that 
the highest star density of our dataset is between 200 and 500 pc (yellow points). The 
top figure shows that only a small set of stars are able to find a planet of the mass 
of Jupiter with the accuracy of DR1 and considering a minimum orbital period of 5 years. 
The middle pannel moves the contour lines to the right increasing the efficiency for the 
detection of objects with $M_{obj} = 10 M_{J}$, and the bottom figure shows the contour 
lines displaced further to the right which shows that sub-stellar objects with 
$M_{obj} = 15 M_{J}$ could be more detectable, however, the higher density of stars 
are beyond our reach to find sub-stellar objects there.}
\label{Figure:Limit-Detection}
\end{figure}

\subsection{Efficiency and effectiveness indicators}
\label{subsec:EEind}

It is worth mentioning that, for multiple sub-stellar objects systems,
the measurements of astrometry correspond to a composition of the
effects produced for each of the objects
\citep[Sect. 5]{Butkevich2017}.  We will assume that each of the stars
houses only one object of mass $M_{obj}$, whose effect on stellar
displacement predominates over the possible effect of others
hypothetical objects in the system \citep[Sect. 1]{Ranalli2017}.  We
define three indicators:

\begin{itemize}
 \item \textbf{Nominal Efficiency (NE)}: We define $N_{nom}$ as the
   number of stars with a minimum detectable period calculated using
   eq. \ref{Equation:Period}, and that contains objects of mass
   $M_{obj}$, and which could be detected with the nominal precision
   of Gaia \citep{Perryman14}. Then, the nominal efficiency will be
   the ratio between $N_{nom}$ and the total number of stars $N$ in
   our data set,

\begin{equation}
\label{Equation:Efficiency-nominal}
NE = \frac{N_{nom}}{N}.
\end{equation}

The minimum detectable periods that we used to calculated $N_{nom}$
are 5 and 10 years respectively (see table
\ref{Table:Period-Min-Allende}).

\item \textbf{DR1 Efficiency (DE)}: This is the ratio between the
  number of stars that contain objects of mass $M_{obj}$, which could
  be detected with the precision of the first release of Gaia (DR1),
  $N_{DR1}$, and the total number of stars in our dataset,

\begin{equation}
\label{Equation:Efficiency-DR1}
DE = \frac{N_{DR1}}{N} \\ \\
\end{equation}

\item \textbf{Effectiveness (Eff)}: It is the ratio between
  \textbf{DE} and \textbf{NE}, and indicates the effectiveness of Gaia
  for the detection of sub-stellar objects. This indicator shows us
  the percentage of success of the mission based on the nominal
  accuracy, and is not affected by the actual existence of objects
  orbiting the stars in the dataset,

\begin{equation}
\label{Equation:Efficiency-DR1}
E_{ff} = \frac{NE}{DE} \\ \\
\end{equation}
\end{itemize}

Both, NE and DE were calculated using Eq. \ref{Equation:Period} for
each object in the dataset, depending on: 1) its respective G-mass, 2)
its distance, 3) its particular nominal precision, which varies
according to the visual magnitude in the G-band, and 4) its particular
observational precision obtained from DR1, which was estimated as the
square root of the sum of the squares of the uncertainties in right
ascension and declination.

\subsection{Effectiveness and efficiency for Allendes' stars}
\label{subsec:Allende}

We applied our effectiveness and efficiency indicators to 12,533 stars
included in the work of \citet{Allende99}, in order to evaluate the
actual accuracy of Gaia using well-known established masses. Table
\ref{Table:Period-Min-Allende} shows that with the nominal precision
indicated by \cite{Perryman14} and considering that the mission time
is 5 years it is possible to find $6,664$ objects with $M_{obj} = 1.0
M_J$ , assuming that in these stars they have planets of the mass of
Jupiter that are responsible for the movement of the star around the
centre of mass. This $6,664$ objects corresponds to 53.172\% of the
stars in the Allendes' catalogue.  With the accuracy of DR1 that
percentage drops to 0\%, and therefore the effectiveness to find
planets of the Jupiter masses is 0\%. In the same way, it is possible
to find $12,532$ objects with $M_{obj} = 10.0 M_J$ that corresponds to
99.992\% of the stars in the Allendes' catalogue (assuming, again, that
these stars have planets with 10 times the mass of Jupiter that are
responsible of the movement of the star around the centre of
mass). With the accuracy of DR1, that percentage drops to 11.7\% and
the corresponding effectiveness is 11.746\%, and so on for the rest of
the objects. 

Down in table \ref{Table:Period-Min-Allende} we show 
the same indicators applied to the $1,477,047$ stars in our dataset. 
We can see that the effectiveness is less than 1\% for all the masses if considering
that the time of operation of Gaia will be 5 years. If we consider
that the operating time extends to 10 years, the effectiveness to find
objects that have 15 times the mass of Jupiter increases to 1.333\%
(that corresponds to $19,191$ objects) and the effectiveness to find
objects that have 20 times the mass of Jupiter increases to 2.421\%
(that corresponds to $35,660$ objects). We applied the effectiveness and 
efficiency, keeping in mind that the mass of
the stars corresponds to our estimation $M_G$, wich according to
Figure \ref{Figure-Error masses} is a good approximation. It should 
be noted that the effectiveness applied to the stars of
Allende is much higher than the effectiveness applied to the stars of
our catalogue because all the stars of Allende are located at a distance
of less than 100 parsec.

\begin{table}
\centering
\scriptsize
\caption{Results for efficiency and effectiveness of DR1-TGAS data, 
  for minimum periods of 5 and 10 years respectively for Allendes' masses (up) and 
  for masses in our dataset (down).}
\vspace{0.4cm}  
\begin{tabular}{ccccccc}
\hline
\hline
{\textbf{Object}} & \textbf{Objs (NE)} & \textbf{NE} & \textbf{Objs (DE)} & \textbf{DE} & \textbf{Eff} & \textbf{Eff Mp} \\
\hline
\multicolumn{6}{l}{\textbf{Allendes' catalogue (12,533 stars)}} \\
\hline
\multicolumn{1}{l}{\textbf{Period 5 years}} \\
\hline    
    1 Jup    & 6664     & 53.172\% & 0        & 0.000\%  & 0.000\%  & 0.000\% \\

    5 Jup    & 12513    & 99.840\% & 1        & 0.008\%  & 0.008\%  & 0.001\% \\

    10 Jup   & 12528    & 99.960\% & 271      & 2.162\%  & 2.163\%  & 0.169\% \\

    15 Jup   & 12532    & 99.992\% & 1215     & 9.694\%  & 9.695\%  & 0.467\% \\

    20 Jup   & 12532    & 99.992\% & 2791     & 22.269\% & 22.271\% & 0.770\% \\
\hline
\multicolumn{1}{l}{\textbf{Period 10 years}} \\
\hline
    1 Jup    & 11799    & 94.143\% & 0        & 0.000\%  & 0.000\%  & 0.000\% \\

    5 Jup    & 12525    & 99.936\% & 84       & 0.670\%  & 0.671\%  & 0.110\% \\

    10 Jup   & 12532    & 99.992\% & 1472     & 11.745\% & 11.746\% & 0.918\% \\

    15 Jup   & 12533    & 100\%    & 4163     & 33.216\% & 33.216\% & 1.598\% \\

    20 Jup   & 12533    & 100\%    & 7026     & 56.060\% & 56.060\% & 1.939\% \\
\hline
\hline
\multicolumn{6}{l}{\textbf{Our dataset (1,477,047 stars)}} \\
\hline
\multicolumn{1}{l}{\textbf{Period 5 years}}\\
\hline
    1 Jup    & 30005    & 2.031\%  & 12       & 0.001\%  & 0.040\%  & 0.020\% \\

    5 Jup    & 552554   & 37.409\% & 595      & 0.040\%  & 0.108\%  & 0.018\% \\

    10 Jup   & 1032432  & 69.898\% & 2880     & 0.195\%  & 0.279\%  & 0.022\% \\

    15 Jup   & 1270598  & 86.023\% & 7125     & 0.482\%  & 0.561\%  & 0.027\% \\

    20 Jup   & 1392728  & 94.291\% & 13212    & 0.894\%  & 0.949\%  & 0.033\% \\
\hline    
\multicolumn{1}{l}{\textbf{Period 10 years}} \\
\hline    
    1 Jup    & 79778    & 5.401\%  & 33       & 0.002\%  & 0.041\%  & 0.020\% \\

    5 Jup    & 876900   & 59.368\% & 1660     & 0.112\%  & 0.189\%  & 0.031\% \\

    10 Jup   & 1298313  & 87.899\% & 8091     & 0.548\%  & 0.623\%  & 0.049\% \\

    15 Jup   & 1439411  & 97.452\% & 19191    & 1.299\%  & 1.333\%  & 0.064\% \\

    20 Jup   & 1472704  & 99.706\% & 35660    & 2.414\%  & 2.421\%  & 0.084\% \\        
\hline   
\end{tabular}
\label{Table:Period-Min-Allende}
\end{table}

\subsection{Effectiveness and efficiency including planetary mass percentiles}
\label{subsec:PMD}

We take into account the bulk distribution of planetary masses
observed. We used 1330 data measurements of $M_P$ taken from the
exoplanet catalogue\footnote{http://exoplanet.eu/}, including
measurements made by radial velocity ($M_P \sin(i)$) and masses
measured directly by other exoplanets detection methods like imaging,
micro lensing, transits and TTV. With this bulk data we calculated
percentiles for planetary masses detectable with Gaia (see table
\ref{Table:PDF-Exoplanets}). We see that 50.977\% of the measured
exoplanetary masses have values less than the $1.0 M_J$ and they are
virtually impossible to detect for Gaia. If we include this
percentiles into the effectiveness indicator, then the probability of
detecting an exoplanet using Gaias' astrometry is reduced. The new
value of the effectiveness is shown in the last column of table
\ref{Table:Period-Min-Allende}. Although the planetary mass distribution function keep being unknown,
observational evidence \citep{Ho11} and statistical approximations 
\citep{Jiang07}, point to the fact that small masses are the most common
among the planets. 

\begin{table}
\centering
\caption{$M_P\sin{i}$ percentiles for exoplanets masses.}
\vspace{0.4cm}
\begin{tabular}{ccc}
\hline
$M_{Obj}$ (Percentile) & Percentage  & Actual detectable fraction \\
\hline
    1 Jupiter       & 50.977\% 	 &  49.023\% \\
    5 Jupiter       & 83.609\%   &  16.391\% \\
    10 Jupiter      & 92.180\%   &  7.820\% \\
    15 Jupiter      & 95.188\%   &  4.812\% \\
    20 Jupiter      & 96.541\%   &  3.459\% \\
\hline
\end{tabular}
\label{Table:PDF-Exoplanets}
\end{table}

\subsection{The top ten objects in DR1-TGAS}
\label{subsec:topten}

Table \ref{Table-Candidates} shows the candidate stars that are
located less than 10 pc and that could contain objects of mass
$M_{obj}$ and with a minimum orbital period $P_{min}$. Here we 
show the best targets to search for exoplanets in DR1-TGAS 
data, according to our efficiency and effectiveness calculations.

\begin{itemize}
\item \textbf{HIP 57367} (GJ 440). On this white dwarf, distanced at
  4.634 pc, is possible the detection of objects starting from a 1
  Saturn mass, with a minimum orbital period of 3.469 year and with
  DR1-TGAS accuracy. This is the best candidate according to our results.
\item \textbf{HIP 82809} (Wolf 629) This is a binary star distanced at
  6.506 pc that may host detectable objects if they have a minimum mass
  of 1 $M_J$.
\item \textbf{HIP 106440} (HD 204961) This high proper-motion star
  hosts two confirmed planets, but a third planet was predicted by
  \cite{Satyal2017}. With the DR1-TGAS accuracy it is possible the 
  detection of objects with masses above 1 $M_J$ and periods of 4.706 yr.
\item \textbf{HIP 93873} (Ross 730). On this high proper-motion star
  it could be detected objects above 1 $M_J$ and with a minimum
  period of 4.793 yr.
\item \textbf{HIP 1475} (GJ 15A) This flare star have a candidate
  companion identified by \cite{Tanner2010} using direct
  imaging. Around this star DR1-TGAS could provide signals only if 
  the object have at least 1 $M_J$ and a period of 3.591 yr.
\item \textbf{HIP 80824} (BD-12 4523) On this BY draconis variable
  star a Jovian exoplanet orbiting could be detected with a
  minimum period of 3.293 yr.
\item \textbf{HIP 23512} (LP 776-46) This high proper-motion star was
  included in the work of \cite{Bozhinova2015}, who studied the
  stellar parameters of the M-dwarf in order to detect low-mass
  planets. With the current Gaias' accuracy it could be detected 
  Jovian planets orbiting this star with a minimum period of 3.781
  yr. Considering that this star is only 9.265 pc away, could be 
  detected planets with masses less than 1 $M_J$ with a slight
  improvement in the accuracy of Gaia-TGAS for nearby stars.
\item \textbf{HIP 91768} (HD 173739) This star is
  located at a distance of 3.527 pc and was included in the work of
  \citep{Leger2015} who did present an analytic model to estimate the
  capabilities of space missions dedicated to the search for
  bio-signatures in the atmospheres of rocky planets located in the
  habitable zone of nearby stars. With the current accuracy it 
  could be detected Jovian planets with a minimum period of 2.972
  yr.
\item \textbf{HIP 29295}(GJ 229) On this flare star could be detected
  Jovian planets of minimum period 3.579 yr. This star was included en
  the work of \cite{Newton2016} who analyzed the impact of stellar
  rotation on the detectability of habitable planets around M-dwarfs.
\item \textbf{HIP 57544}(GJ 445) This is another flare star, distanced 
  at 5.225 pc. It is possible to detect objects starting from a 1 $M_J$, 
  with a minimum orbital period of 4.121 yr.
\end{itemize}

With this analysis we are not ensuring that sub-stellar objects will be 
found around these stars, but we believe that given Gaias' observational 
capability, these would be the best candidates to search among the astrometry 
data. We know that as the mission releases new results, it is expected that 
its accuracy will improve, which will facilitate the search for objects. We 
believe that some stars on our list could be feasible for analysis in 
search of exoplanets.

\begin{table}
\centering
\caption{List of Gaias' stars that are located less than 10 pc with alleged objects 
of mass $M_{obj}$ and with a measurable minimum orbital period $P_{min}$.}
\vspace{0.4cm}
\tiny
\begin{tabular}{ccccrcccc}
\hline
\hline
\textbf{Id} & \textbf{Distance (pc)} & \textbf{$M_{obj}$} & \textbf{$P_{min}$ (yr)} & & \textbf{Id} & \textbf{Distance (pc)} & \textbf{$M_{obj}$} & \textbf{$P_{min}$ (yr)} \\
\hline
    HIP 101180 & 8.056    & 10 Jupiter & 0.433    & & HIP 53020 & 6.966    & 10 Jupiter & 0.520 \\
             & & 15 Jupiter & 0.236    & &          & & 15 Jupiter & 0.283 \\
    HIP 102409 & 9.792    & 10 Jupiter & 0.574    & & HIP 55360 & 9.120    & 10 Jupiter & 1.346 \\
             & & 15 Jupiter & 0.312    & &          & & 15 Jupiter & 0.733 \\
    HIP 103096 & 7.034    & 10 Jupiter & 0.442    & & HIP 56452 & 9.561    & 10 Jupiter & 1.013 \\
             & & 15 Jupiter & 0.241    & &          & & 15 Jupiter & 0.551 \\
    HIP 105090 & 3.982    & 10 Jupiter & 0.256    & & HIP 56528 & 9.126    & 10 Jupiter & 0.346 \\
             & & 15 Jupiter & 0.140    & &          & & 15 Jupiter & 0.188 \\
    HIP 106440 & 4.972    & 1 Jupiter & 4.706    & & HIP 57087 & 9.749    & 10 Jupiter & 0.292 \\
             & & 10 Jupiter & 0.149    & &          & & 15 Jupiter & 0.159 \\
             & & 15 Jupiter & 0.081    & & HIP 57367 & 4.634    & 1 Saturn & 3.469 \\
    HIP 109388 & 8.830    & 10 Jupiter & 0.496    & &          & & 1 Jupiter & 0.568 \\
             & & 15 Jupiter & 0.270    & &          & & 10 Jupiter & 0.018 \\
    HIP 111802 & 8.875    & 10 Jupiter & 0.608    & &          & & 15 Jupiter & 0.010 \\
             & & 15 Jupiter & 0.331    & & HIP 57544 & 5.225    & 1 Jupiter & 4.121 \\
    HIP 113020 & 4.672    & 10 Jupiter & 0.458    & &          & & 10 Jupiter & 0.130 \\
             & & 15 Jupiter & 0.249    & &          & & 15 Jupiter & 0.071 \\
    HIP 113229 & 8.605    & 10 Jupiter & 0.282    & & HIP 57548 & 3.381    & 1 Jupiter & 1.571 \\
             & & 15 Jupiter & 0.153    & &          & & 10 Jupiter & 0.050 \\
    HIP 113296 & 6.850    & 10 Jupiter & 0.513    & &          & & 15 Jupiter & 0.027 \\
             & & 15 Jupiter & 0.280    & & HIP 57802 & 8.773    & 10 Jupiter & 0.202 \\
    HIP 113576 & 8.194    & 10 Jupiter & 0.476    & &          & & 15 Jupiter & 0.110 \\
             & & 15 Jupiter & 0.259    & & HIP 62452 & 8.056    & 10 Jupiter & 0.398 \\
    HIP 117473 & 5.914    & 10 Jupiter & 0.193    & &          & & 15 Jupiter & 0.216 \\
             & & 15 Jupiter & 0.105    & & HIP 66906 & 9.116    & 10 Jupiter & 0.311 \\
    HIP 120005 & 6.291    & 10 Jupiter & 0.359    & &          & & 15 Jupiter & 0.169 \\
             & & 15 Jupiter & 0.195    & & HIP 71253 & 6.284    & 10 Jupiter & 0.399 \\
    HIP 14101 & 6.734    & 10 Jupiter & 0.664    & &          & & 15 Jupiter & 0.217 \\
             & & 15 Jupiter & 0.362    & & HIP 74995 & 6.303    & 1 Jupiter & 4.547 \\
    HIP 1475 & 3.562    & 1 Jupiter & 3.591    & &          & & 10 Jupiter & 0.144 \\
             & & 10 Jupiter & 0.114    & &          & & 15 Jupiter & 0.078 \\
             & & 15 Jupiter & 0.062    & & HIP 76074 & 5.926    & 10 Jupiter & 0.237 \\
    HIP 21088 & 5.509    & 10 Jupiter & 0.269    & &          & & 15 Jupiter & 0.129 \\
             & & 15 Jupiter & 0.146    & & HIP 7751 & 8.097    & 10 Jupiter & 1.370 \\
    HIP 21553 & 9.878    & 10 Jupiter & 2.181    & &          & & 15 Jupiter & 0.746 \\
             & & 15 Jupiter & 1.187    & & HIP 80459 & 6.490    & 10 Jupiter & 0.500 \\
    HIP 21932 & 9.405    & 10 Jupiter & 0.920    & &          & & 15 Jupiter & 0.272 \\
             & & 15 Jupiter & 0.501    & & HIP 80824 & 4.305    & 1 Jupiter & 3.293 \\
    HIP 23512 & 9.265    & 1 Jupiter & 3.781    & &          & & 10 Jupiter & 0.104 \\
             & & 10 Jupiter & 0.120    & &          & & 15 Jupiter & 0.057 \\
             & & 15 Jupiter & 0.065    & & HIP 82003 & 9.844    & 10 Jupiter & 0.399 \\
    HIP 23932 & 9.357    & 10 Jupiter & 0.753    & &          & & 15 Jupiter & 0.217 \\
             & & 15 Jupiter & 0.410    & & HIP 82809 & 6.506    & 1 Jupiter & 2.459 \\
    HIP 25878 & 5.647    & 10 Jupiter & 0.721    & &          & & 10 Jupiter & 0.078 \\
             & & 15 Jupiter & 0.392    & &          & & 15 Jupiter & 0.042 \\
    HIP 29277 & 9.383    & 10 Jupiter & 0.507    & & HIP 85295 & 7.736    & 10 Jupiter & 0.270 \\
             & & 15 Jupiter & 0.276    & &          & & 15 Jupiter & 0.147 \\
    HIP 29295 & 5.792    & 1 Jupiter & 3.579    & & HIP 86057 & 9.925    & 10 Jupiter & 2.148 \\
             & & 10 Jupiter & 0.113    & &          & & 15 Jupiter & 1.169 \\
             & & 15 Jupiter & 0.062    & & HIP 86162 & 4.545    & 10 Jupiter & 0.217 \\
    HIP 31292 & 8.832    & 10 Jupiter & 0.826    & &          & & 15 Jupiter & 0.118 \\
             & & 15 Jupiter & 0.449    & & HIP 91768 & 3.527    & 1 Jupiter & 2.972 \\
    HIP 31293 & 8.843    & 10 Jupiter & 0.355    & &          & & 10 Jupiter & 0.094 \\
             & & 15 Jupiter & 0.193    & &          & & 15 Jupiter & 0.051 \\
    HIP 33226 & 5.546    & 10 Jupiter & 0.462    & & HIP 91772 & 3.492    & 10 Jupiter & 0.195 \\
             & & 15 Jupiter & 0.252    & &          & & 15 Jupiter & 0.106 \\
    HIP 33499 & 7.899    & 10 Jupiter & 0.608    & & HIP 93873 & 8.815    & 1 Jupiter & 4.793 \\
             & & 15 Jupiter & 0.331    & &          & & 10 Jupiter & 0.152 \\
    HIP 40501 & 8.926    & 10 Jupiter & 1.178    & &          & & 15 Jupiter & 0.083 \\
             & & 15 Jupiter & 0.641    & & HIP 93899 & 8.867    & 10 Jupiter & 0.218 \\
    HIP 46655 & 9.998    & 10 Jupiter & 0.608    & &          & & 15 Jupiter & 0.119 \\
             & & 15 Jupiter & 0.331    & & HIP 94761 & 5.902    & 10 Jupiter & 0.180 \\
    HIP 47103 & 9.399    & 10 Jupiter & 0.928    & &          & & 15 Jupiter & 0.098 \\
             & & 15 Jupiter & 0.505    & & HIP 96100 & 5.760    & 10 Jupiter & 2.438 \\
    HIP 47425 & 9.615    & 10 Jupiter & 0.402    & &          & & 15 Jupiter & 1.327 \\
             & & 15 Jupiter & 0.219    & & HIP 9786 & 9.323    & 10 Jupiter & 0.331 \\
    HIP 47780 & 9.450    & 10 Jupiter & 0.900    & &          & & 15 Jupiter & 0.180 \\
             & & 15 Jupiter & 0.490    & & HIP 99701 & 6.164    & 10 Jupiter & 0.394 \\
    HIP 4856 & 8.240    & 10 Jupiter & 0.335    & &          & & 15 Jupiter & 0.215 \\
             & & 15 Jupiter & 0.183    & & TYC 3980-1081-1 & 8.292    & 10 Jupiter & 1.306 \\
    HIP 4872 & 9.849    & 10 Jupiter & 0.589    & &          & & 15 Jupiter & 0.711 \\
             & & 15 Jupiter & 0.320    & &          & &          &  \\
\hline
\end{tabular}
\normalsize
\label{Table-Candidates}
\end{table}

\section{Discussion and conclusions}
\label{sec:DisCon}

The precision of DR1-TGAS is $\sim 200 \mu as$ for right ascension and
declination, and $\sim 280 \mu as$ for parallax respectively. The
astrometric angle that describe a star at 100 pc of distance moving
around the centre of mass of the system, and with a period of 5 years,
indicates that the accuracy of DR1 is not enough to detect objects of
mass $M_{obj} = 1.0 \ M_J$ (see table \ref{Table-Parameters}).  An
object of mass $M_{obj} = 10 \ M_J$ will be detectable if it orbits
stars of $M_\star < 0.4 \ M_\odot$. An object of $20 \ M_J$ will be
detectable if it orbits stars of less than $1.0 \ M_\odot$. That is,
the accuracy of DR1-TGAS only allows the detection of giant sub-stellar
objects that orbit low mass stars.

Although you must be aware of the preliminary character of this first
release, it is undoubted that the precision of the measures faces a
great challenge, as they must be improved by a factor of 10 for the
search of Jovian planets in a radius of 100 pc and periods of 5 years
(see Table \ref{Table-Parameters}), if you go up to 500 pc, the factor
increases up to 50.

According to our results in table \ref{Table:Period-Min-Allende}, and
for a minimum detectable period of 10 years, if we consider a mission
duration of 10 years, we could find sub-stellar objects of mass
$M_{obj} = 10 \ M_J$ in the $88\%$ of the stars of our dataset, this
is, assuming, of course, that all the stars in our dataset could host
objects with this mass, and that the precision in the measurements of
position is equal to the nominal precision. However, with the accuracy
of DR1, the percentage of sub-stellar objects of mass $M_{obj} = 10
\ M_J$ that we would find is reduced to only $0.5\%$. It should be noted 
that the use of Gaias' astrometry for the search of
sub-stellar objects requires that all observations made during the
mission have, at least, the same minimum required precision.

Therefore, if Gaias' operating time increases to 10 years, and the
nominal precision in the position measurements is achieved in the
fifth year of operation, then we would have 5 years of observations
with the nominal precision, and according to table
\ref{Table:Period-Min-Allende}, for a minimum period of 5 years,
we could find sub-stellar objects of mass $M_{obj} = 10 \ M_J$ in the
$70\%$ of the stars of our dataset, again, assuming that all the stars
of our dataset host objects of mass $M_{obj} = 10 \ M_J$.

The present analysis shows us the magnitude of the challenge that is
assumed for the Gaia project in order to minimize uncertainty in the
measurements of the position of the stars, since the precision of the
data in the DR1-TGAS only offer possibility of detecting a few massive
objects with periods above 5 years and that orbits dwarf stars at a
distance of up to 100 pc.

\section*{Acknowledgements}

We thank the referee for the valuable insights and comments, all of
them have been included in the final version of the text.  FACom -
SEAP group is supported by \textit{Estrategia de Sostenibilidad
  2016-2017}, Vicerector\'ia de Investigaci\'on - UdeA. This work 
  has made use of data from the European Space Agency (ESA)
mission Gaia (https://www.cosmos.esa.int/gaia), processed
by the Gaia Data Processing and Analysis Consortium (DPAC, 
https://www.cosmos.esa.int/web/gaia/dpac/consortium). Funding
for the DPAC has been provided by national institutions, in particular
the institutions participating in the Gaia Multilateral
Agreement.

\section*{References}

\bibliographystyle{elsarticle-harv}
\bibliography{bibliography}

\end{document}